\documentclass[final,5p,times,twocolumn]{elsarticle}

\usepackage{graphicx}

\journal{Nuclear Instruments Method}
\begin{document}
\begin{frontmatter}
\title{Distillation of hydrogen isotopes for polarized HD targets}
\author[a]{T.~Ohta\corref{cor1}}
\cortext[cor1]{Corresponding author.}
\ead{takeshi@rcnp.osaka-u.ac.jp}

\author{S.~Bouchigny$^{b,c}$, J.-P.~Didelez$^{b}$, M.~Fujiwara$^{a}$, K.~Fukuda$^{d}$, H.~Kohri$^{a}$, T.~Kunimatsu$^{a}$, C.~Morisaki$^{a}$, S.~Ono$^{a}$, G.~Rouill\'e$^{b}$, M.~Tanaka$^{e}$, K.~Ueda$^{a}$, M.~Uraki$^{a}$, M.~Utsuro$^{a}$, S.Y.~Wang$^{f,g}$, and M.~Yosoi$^{a}$}
\address[a]{ Research Center for Nuclear Physics, Osaka University, Mihogaoka 10-1, Ibaraki, Osaka 567-0047, Japan}
\address[b]{IN2P3, Institut de Physique Nucl\'{e}aire, F-91406 ORSAY, France}
\address[c]{CEA LIST, BP6-92265 Fontenay-aux-Roses, CEDEX, France}
\address[d]{Kansai University of Nursing and Health Sciences, Shizuki Awaji 656-2131, Japan}
\address[e]{Kobe Tokiwa University, Ohtani-cho 2-6-2, Nagata, Kobe 653-0838, Japan}
\address[f]{Institute of Physics, Academia Sinica, Taipei 11529, Taiwan}
\address[g]{Department of Physics, National Kaohsiung Normal University, Kaohsiung 824, Taiwan}

\begin{abstract}
We have developed a new cryogenic distillation system to purify
Hydrogen-Deuteride (HD) gas for polarized HD targets in LEPS experiments at SPring-8.
A small amount of ortho-H$_2$ ($\sim$0.01\%) in the HD gas plays an important role in efficiently polarizing the HD target.
Since there are 1$\sim$5\% impurities of H$_2$ and D$_2$ in commercially available HD gases, 
it is necessary to purify the HD gas up to $\sim$99.99\%.
The distillation system is equipped with a cryogenic distillation unit filled with many small stainless steel cells called "Heli-pack".
The distillation unit consists of a condenser part, a rectification part, and a reboiler part. 
The unit is kept at the temperature of 17$\sim$21 K.
The Heli-pack has a large surface area that makes a good contact between gases and liquids.
An amount of 5.2 mol of commercial HD gas is fed into the distillation unit.
Three trials were carried out to purify the HD gas by changing temperatures (17.5 K and 20.5 K) and gas extraction speeds (1.3 ml/min and 5.2 ml/min).
The extracted gas was analyzed using a gas analyzer system combining a quadrupole mass spectrometer with a gas chromatograph. 
One mol of HD gas with a purity better than 99.99\% has been successfully obtained for the first time. 
The effective NTP (Number of Theoretical Plates), which is an indication of the distillation performances, is obtained to be 37.2$\pm$0.6. 
This value is in good agreement with a designed value of 37.9.
The HD target is expected to be efficiently polarized under a well-controlled condition by adding an optimal amount of ortho-H$_2$ to the purified HD gas.
 
\end{abstract}
\end{frontmatter}

\section{Introduction}
\label{intro}
Hadron photoproduction experiments using polarized photon beams and polarized nucleon targets are of importance to study the hidden $s\bar{s}$-quark content of the nucleon structure~\cite{Titov97}, the reaction mechanisms of $\phi$ and $K$ meson photoproduction ~\cite{Mibe05,Kohri06,Kohri10,Sandorfi10}, and exotic hadron properties~\cite{Nakano09}.
In order to measure double polarization observables at SPring-8, we are developing technologies to produce a frozen-spin polarized Hydrogen-Deuteride (HD) target~\cite{Fujiwara03}.

The HD molecule was proposed as a frozen-spin polarized target in Refs.~\cite{Bloom57,Hardy66,Honig67} by using the mechanism clarified by Motizuki et al.~\cite{T. Moriya,K. Motizuki}.
A small amount ($\sim$0.01\%) of ortho-H$_2$ (o-H$_2$) with a spin of J=1 is mixed with the pure HD gas. 
This HD gas is solidified and is kept for a long time at a temperature lower than 14 mK and in a high magnetic field of 17 T.
The spin-flip process between the HD and the o-H$_2$ gradually grows the HD polarization.
The o-H$_2$ is converted to the para-H$_2$ (p-H$_2$) with a spin of J=0 after a long period of cooling.
When all the o-H$_2$ molecules are converted to the p-H$_2$, the relaxation time of hydrogen polarization in the HD target becomes very long even at a temperature of $\sim$4 K and in a holding field around 1T.
The polarization degree of hydrogen is expected to be higher than 84\% and is measured by using the NMR method~\cite{PXI-NMR}.
Many trials have been exerted to realize the polarized HD target at Syracuse \cite{Honig89,Honig95}, BNL \cite{Wei00,Wei01,Wei04}, and ORSAY \cite{Breuer98,Rouille01,Bassan04,Bouchigny05,Bouchigny09,BouchignyThesis}.
Recently, the HD target was used in the LEGS experiment~\cite{Holbit09} for the first time, and will be used at JLab \cite{Sandorfi} and at SPring-8 \cite{Kohri10-1} in the near future.

One of the common problems is related to the measurement of the HD gas concentration purified by distillation.
In the past, the purity of the HD gas was measured by using a quadrupole mass spectrometer (QMS).
The HD gas is ionized in the QMS by electron bombardment, which produces not only HD$^+$ but also D$^+$.
The D$^+$ is mis-identified as the H$_{2}^{+}$ impurity because the mass and charge are the same.
Therefore, the amounts of impurities were not precisely determined, and the distillation performance was not checked correctly.
However, the afore-mentioned problem was solved by introducing a new gas analyzer system combining a gas chromatograph with the QMS~\cite{GC-QMS}.
This new gas analyzer enables us to measure the concentrations of the impurities with a precision better than 0.01\%.

Since the commercial HD gas has about 1$\sim$5\% impurities of H$_2$ and D$_2$, we need to purify it up to $\sim$99.99\% by distillation.
After the distillation, an optimal amount of the o-H$_2$ is added to efficiently polarize the HD target.
In the past, we used a cryogenic distillation system, which had a rectification column containing about 20 cells called Stedman packing~\cite{Bouchigny09,BouchignyThesis}, provided by the ORSAY group in 2006.
This distillation system required us to prepare the subsequent second distillation in order to obtain a high purity HD gas for the polarized HD target.
The HD gas was distilled, and the H$_2$ component was removed in the first distillation.
Next the HD gas was again distilled, and the D$_2$ component was removed in the second distillation.
The periods of the first and second distillations were 20 days and 10 days, respectively.
Another group used a distillation system consisting of many stainless coiled ribbons, called Dixon Ring, for separating hydrogen isotopes~\cite{M. Enoeda}.
The Dixon Ring has a large surface area, giving a good separation of the hydrogen isotopes.

We have developed a new cryogenic distillation system containing many small stainless steel cells called "Heli-pack"~\cite{to-toku}.
The Heli-pack, which has a surface area larger than the Dixon Ring, is expected to give a better separation of the hydrogen isotopes.
In this paper, we report the performance of the newly developed cryogenic distillation system for producing the pure HD gas.

\section{Experiment}
\subsection{Principle}
In general, the distillation system is used to separate a desired chemical component from the mixed compound by utilizing the vaporability difference.
A low-boiling component vaporizes, and is separated from others as a gas. Then high-boiling components are liquefied. 
In chemical industries, this kind of the distillation method is used to acquire a high purity chemical component for large-volume production. 

In the case of the HD distillation, H$_2$, HD, and D$_2$ gases are liquefied by a refrigerator at a low temperature.
Using differences between the boiling points among H$_2$, HD, and D$_2$, HD gas is separated from the others.
At the beginning of the distillation process, the low-boiling component, H$_2$, is kept as a gas at the upper part of the distillation unit.
The other (HD and D$_{2}$) components drop down to the bottom of the distillation unit as liquids.
Packs, which promote the separation of the components, are filled in the distillation unit and make a good contact between gases and liquids. The bottom of the distillation unit is warmed up by radiation heating. 
Since the middle part of the distillation unit is made of stainless steel with small heat conduction, a temperature gradient is realized.
The H$_2$ gas is extracted from the top through a pipe connected to a gas storage tank. 
Then, the remaining HD component vaporizes with a high purity after extracting the H$_2$ component. The HD is extracted to another tank, and is used for the polarized HD target.

\begin{figure}[h!]
  \begin{center}
   \includegraphics[width=85mm]{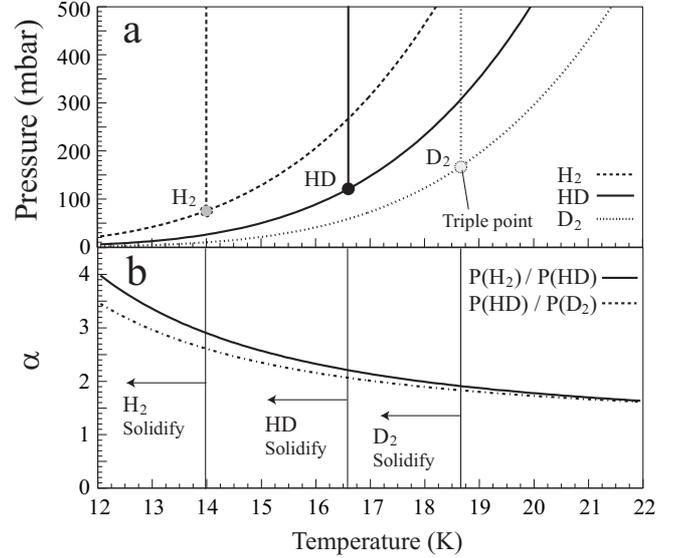}
  \end{center}
  \caption{(a) Phase diagram of H$_2$, HD, and D$_2$. Triple points for three components are indicated by the circles. The lines in the right side of the triple points show the border of the transition between liquid and gas (vapor pressure curve). The vertical lines show the border of the transition between solid and liquid (melting curve). The lines in the left side of the triple points show the border of the transition between solid and gas (sublimation curve).
These curves connect at the triple point.
(b) Relative volatility $\alpha$ for P(H$_{2}$)/P(HD) and P(HD)/P(D$_{2}$) as a function of temperature. 
}  
\label{fig:VaporCurve.eps}
\end{figure}

As seen in Fig.~\ref{fig:VaporCurve.eps}(a)~\cite{Chem. Eng. Hand.}, the H$_2$ component is easily evaporated in comparison with the HD component at 17 K.
The vaporability of a low-boiling component to a high-boiling component is expressed by using the vapor pressure ratio of the two components, and is defined in terms of the relative volatility $\alpha$ as a function of temperature.
Fig.~\ref{fig:VaporCurve.eps}(b) shows a relative volatility $\alpha$ for P(H$_{2}$)/P(HD) and P(HD)/P(D$_{2}$).
The relative volatility increases with decreasing the temperature.
The distillation efficiency increases with increasing relative volatility.
However, the distillation efficiency decreases when HD is solidified.
Therefore, the distillation efficiency is expected to be optimum at around 17 K.

The separation ability of the distillation is expressed in terms of NTP (Number of Theoretical Plates), which is defined by using the concentrations of a low-boiling component at the bottom and upper parts of the distillation unit~\cite{M. R. Fenske}.
In the case of the distillation of the HD gas, the NTP is given as;
\begin{equation}
NTP= \frac{ \displaystyle \ln \left(\frac{[H_2]_{Top}}{1-[H_2]_{Top}}\right) - \ln \left( \frac{[H_2]_{Bot}}{1-[H_2]_{Bot}}\right) }{ \ln \alpha},
\label{equ:NTP}
\end{equation}
where [H$_2$]$_{Top}$ and [H$_2$]$_{Bot}$ are the concentrations of the H$_2$ gas at the top and bottom parts, respectively, and $\alpha$ is the relative volatility between H$_2$ and HD.
To increase the separation ability, it is important to increase the NTP.
The NTP depends on the structure and the surface area of the packs and the height of the stacked column.
We designed the distillation system with NTP=37.9 as discussed later.

\subsection{Apparatus}
\begin{figure}[htbp]
  \begin{center}
    \includegraphics[width=80mm]{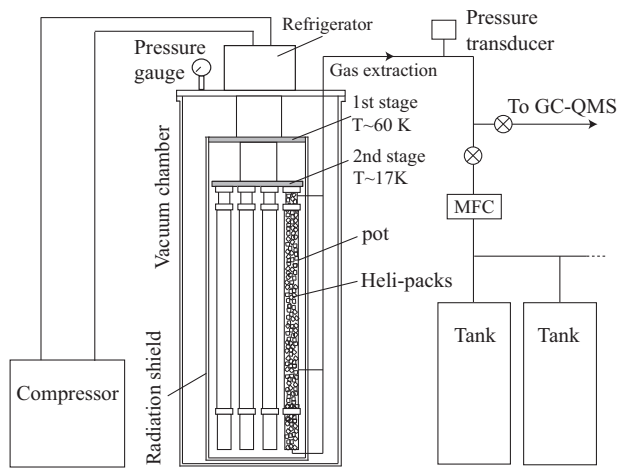}
  \end{center}
  \caption{Schematic view of the HD gas distillation system. The distillation system is equipped with the refrigerator to liquefy the commercial HD gas. MFC is a mass flow controller. The mass flow controller (MC-3000L) is made by LINTEC Corporation. A pressure transducer is installed at the gas extraction line. GC-QMS is a gas analyzer system. The radiation shield is made of copper with a thickness of 1 mm. } 
  \label{fig:Distiller_HELIPACK.eps}
\end{figure}

Fig.~\ref{fig:Distiller_HELIPACK.eps} shows a schematic view of the HD gas distillation system. 
The upper part of the distillation unit is cooled with a cryogenic panel directly connected to the refrigerator, and the bottom part is warmed by radiation heating.
The radiation shield is connected to the first stage of the refrigerator and is kept at a temperature of about 90$\sim$100 K.
The distillation unit is cooled down to around 17 K.
We use the refrigerator system (Refrigerator unit: RDK-408S, Compressor unit: CSW-71C) produced by Sumitomo Heavy Industries.
The cooling power of the refrigerator is 35 W at 45 K in the first stage and 6.3 W at 10 K in the second stage. 
The lowest temperature of the second stage is 7 K. 
The temperatures and pressure in the distillation unit were monitored periodically.
The distillation system has additional three distillation units to increase the productivity of pure HD gas in the future. Only one distillation unit was used in this experiment.
The stainless steel packing called "Heli-pack" is filled into the distillation unit to promote the separation among the H$_2$, HD, and D$_2$ components.
Heat exchange between gas and liquid takes place on the surface of the Heli-pack cells. 
A gas with a low-boiling component is extracted from the upper part of the distillation unit to the tanks through the mass flow controller.

The impurity of the extracted gas is monitored with a hydrogen isotope analyzer system which combines the Gas Chromatograph with the Quadrupole Mass Spectrometer (GC-QMS) ~\cite{GC-QMS}.
In the gas chromatograph, a narrow capillary tube with a diameter of 0.50 mm is cooled at 110 K. 
Neon gas is used as a carrier gas.
The p-H$_2$, o-H$_2$, HD, and D$_2$ are separated using the retention time of the narrow capillary tube and injected to the QMS.
In the QMS, the analyzed gas is ionized at an ion source by electron bombardment.
Ions are mass-separated according to the mass/charge ratio (u/e). 
The GC-QMS enables us to observe the p-H$_2$, o-H$_2$, HD, and D$_2$ gases separately by measuring the retention time in the GC and by determining the mass/charge ratios in the QMS.

\begin{figure}[htbp]
  \begin{center}
    \includegraphics[width=70mm]{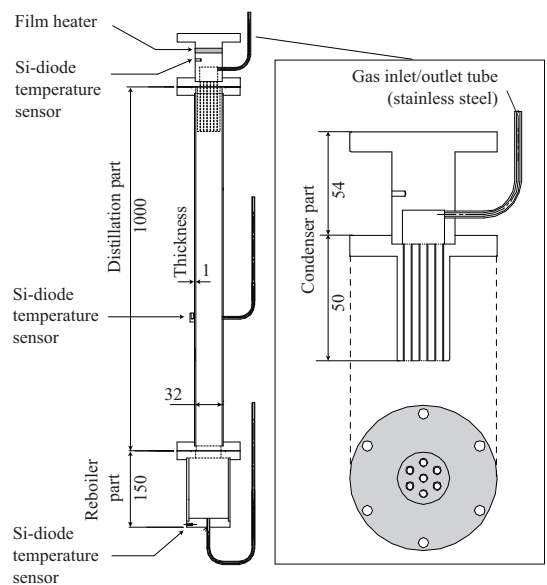}
  \end{center}
  \caption{Sectional drawing of the distillation unit. At the bottom of the condenser part, seven screw holes are prepared for solidifying the commercial HD gas. All the numerical numbers are given in unit of mm.}
  \label{fig:pot.eps}
\end{figure}

 Fig.~\ref{fig:pot.eps} shows a sectional drawing of the distillation unit.
The distillation unit consists of three parts. Each part has a gas inlet/outlet tube.
The condenser part is made of copper. 
The outer casing of the rectification part is made of stainless steel.
The outer casing of the reboiler part is made of copper.
The specification of the distillation unit is listed in Table \ref{table:Spec}.
Three silicon-diode temperature sensors are set for monitoring the temperature of the condenser, rectification, and reboiler parts.
A film heater for controlling the temperature is wound around the condenser part. 
The power of the heater is 20$\sim$30 W. 
The reboiler part is warmed up by radiation heating.
At the condenser part, the H$_2$ gas can only pass through to the extraction pipe, and the HD and D$_2$ are liquefied. 

\begin{table}[htbp]
 \caption{Specifications of the distillation unit and Heli-pack. The notation, SWG, indicates Standard Wire Gauge. HETP (Height Equivalent to a Theoretical Plate)~\cite{to-toku} indicates the Heli-pack separation ability.}
\begin{center}
  \small{ 
  \begin{tabular}{ll}

     \hline
    \textbf {Specification of the distillation unit}&\\    
Material and Height       &    \\
\quad Condenser part & Copper 110 mm\\
\quad Rectification part & Stainless steel 1000 mm\\
\quad Reboiler part &   Copper 150 mm\\
Inner diameter &    32 mm\\
Inner volume &   1 L \\
Packed length &   1100 mm \\
  \hline
  \textbf{Specification of the Heli-pack~\cite{to-toku}}& \\
HETP & 29 mm \\
Wire diameter (SWG) &    0.0193 mm (\#36)\\
Surface area &    3160 m$^2$/m$^3$\\
Free volume &    97.1\%\\
Density & 1450 kg/m$^3$    \\
Material & Stainless steel \\
    \hline
  \end{tabular}
  }
 \end{center}
 \label{table:Spec}
\end{table}

\begin{figure}[h]
  \begin{center}
    \includegraphics[width=50mm]{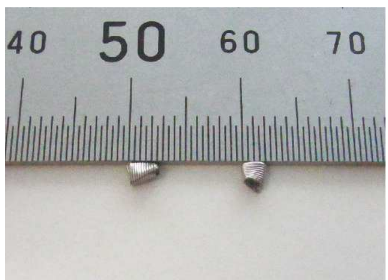}
  \end{center}
  \caption{Photograph of Heli-packs which are filled into the rectification and reboiler parts.}
  \label{fig:HELIPACK.eps}
\end{figure}

Fig.~\ref{fig:HELIPACK.eps} shows a photograph of the Heli-packs. 
The distillation unit is filled with about 100,000 Heli-packs. Each pack has a rolled shape like a coil and is twisted for efficient heat conduction between gas and liquid. 
The NTP is determined from the HETP of the Heli-pack theoretically calculated as 
\begin{equation}
NTP= \frac{Packed\ length}{HETP}=\frac{1100\ mm}{29 \ mm}=37.9
\end{equation}
With NTP=37.9, the H$_2$/HD ratio is enhanced by a factor of 1.74$^{37.9}$ when the relative volatility $\alpha$ is assumed to be 1.74 at 20.5 K.
If the H$_2$ concentration is 0.001\% at the reboiler part, the H$_2$ is purified to 99.993\% at the condenser part, which is derived from Equation (\ref{equ:NTP}). 

\subsection{Distillation procedure}
We operated three times to obtain pure HD gases by changing experimental parameters. 
Table~\ref{fig:conditions.table} lists the experimental parameters for the distillation.
The amount of 5.2 mol of commercial HD gases in the tanks were fed into the distillation unit through the inlet/outlet tube, and solidified at the condenser part which was cooled at 8 K.
The valve between the distillation unit and the tanks was closed after solidifying the HD gases. 
The HD solid was melted by heating up and was liquefied.
It took 6 hours until realizing the equilibrium of H$_2$, HD, and D$_2$.
The concentration of H$_2$ increased to more than 99\%.
The gas was extracted from the condenser part to the gas storage tanks made of stainless steel through the mass flow controller with a constant flow rate.
The concentrations of H$_2$, HD, and D$_2$ were monitored by the GC-QMS during the gas extraction process. The temperature of the condenser part was controlled in the range of 17$\sim$21 K by changing the heater power.
\begin{table}[htbp]
 \caption{Experimental parameters for the distillation.}
 \begin{center}
  \begin{tabular}{lccc}
    \hline
     Condition  &  Run 1 &  Run 2 &  Run 3  \\
    \hline
    Temperature (K) & 17.5 & 20.5 & 20.5 \\
       Extraction speed (ml/min) &  1.3 & 1.3 & 5.2\\
        Experimental period (day) & 17 & 17 & 7 \\
    \hline
  \end{tabular}
 \end{center}
 \label{fig:conditions.table}
\end{table}

\section{Experimental results and analysis}
\subsection{Gas analysis of commercial HD gas}
\begin{figure}[htbp]
  \begin{center}
    \includegraphics[width=85mm]{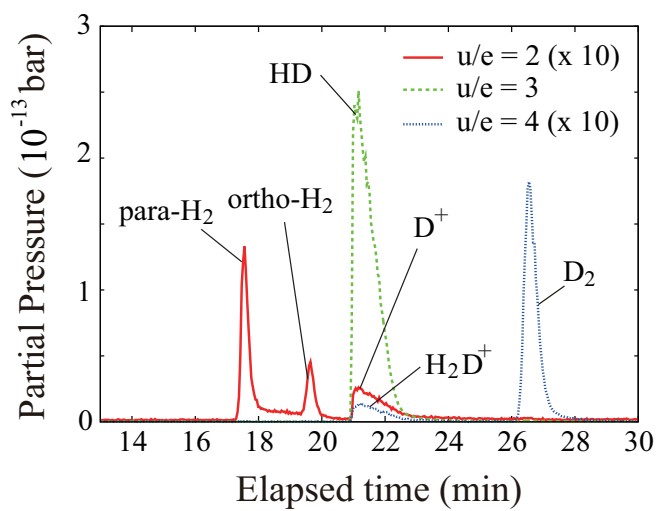}
  \end{center}
  \caption{Results of the gas analysis with the GC-QMS for the commercial HD gas. The horizontal axis is the elapsed time after the gas injection to the gas chromatograph. The vertical axis is the partial pressure of each gas measured with the QMS. u/e is the mass/charge ratio.
The solid line, dashed line, and dotted line show the spectra with u/e=2, 3, and 4, respectively.
}  \label{fig:concentrations_ini.eps}
\end{figure}

Fig.~\ref{fig:concentrations_ini.eps} shows the results of the gas analysis for a commercial HD gas. 
Peaks of p-H$_2$, o-H$_2$, HD, and D$_2$ are observed with a good resolution. 
The continuous component between the p-H$_2$ and o-H$_2$ peaks is caused from the para-ortho transition of H$_2$ in the capillary tube.
The H$_2$, HD, and D$_2$ yields have been obtained by integrating the peak areas.
The concentrations are calibrated by dividing the obtained yields by factors of 2.18 for H$_2$ and of 0.77 for D$_2$ because the sensitivities are different from those for the HD component in the QMS measurement.
These factors are the relative sensitivities to the HD, and are determined by analyzing calibration gases.
The calibrated concentrations of the commercial HD gas are 1.327$\pm$0.003\%, 93.127$\pm$0.008\%, and 5.546$\pm$0.007\%, respectively for H$_2$, HD, and D$_2$. 

The calibration gases have been prepared by mixing the HD gas and another gas (H$_2$ or D$_2$) with a ratio of 1:1.
The factor of 2.18 is obtained from the analysis of a gas with HD and H$_2$.
The factor of 0.77 is obtained from the analysis of a gas with HD and D$_2$.
At the same location of the HD peak, other peaks are observed in the spectra with u/e=2 and u/e=4.
These peaks are due to D$^+$ and H$_2$D$^+$, respectively, produced from the HD component as fragmentations at the ionization process in the QMS. 
A shape of the peak depends on the amounts of the components in the injected gas. 
When the amounts in the injected gas are relatively smaller than a receptible volume of the capillary tube, the width of the peaks is narrow.
On the other hand, when the amounts are much larger than the volume of the capillary tube, we observe a broad peak.
However, it is found that the distortion of the peak shape does not give any serious effect in determining the gas concentrations~\cite{GC-QMS}.

\subsection{Gas analysis of pre-extraction and effective NTP}
\begin{figure}[htbp]
  \begin{center}
    \includegraphics[width=85mm]{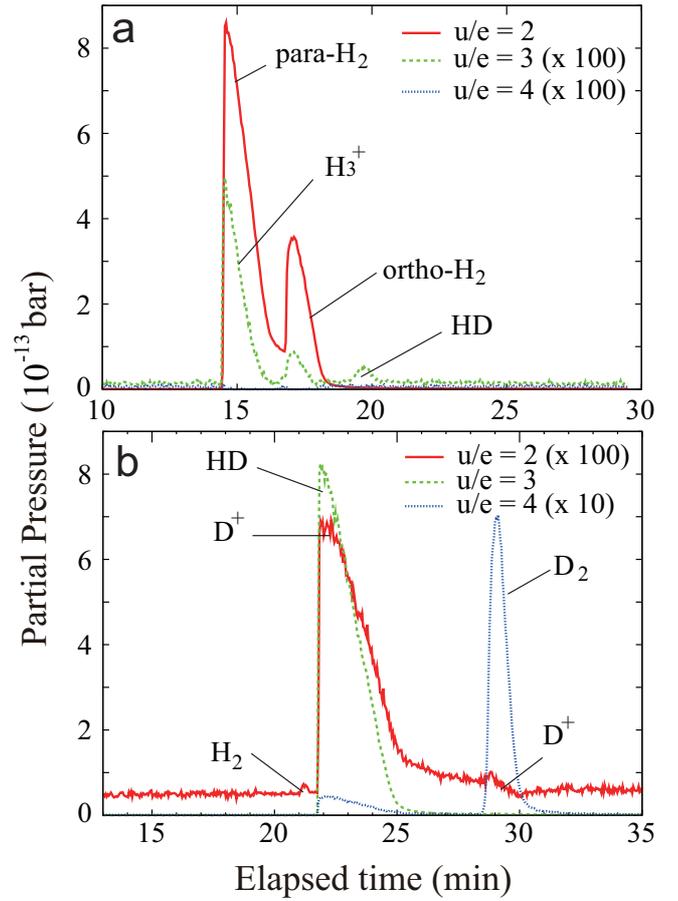}
  \end{center}
  \caption{Results of the gas analysis with the GC-QMS for the gas from the condenser (a) and reboiler (b) parts before the gas extraction operation. Notations are the same as in Fig.~\ref{fig:concentrations_ini.eps}. }
\label{fig:concentrations_before.eps}
\end{figure}

After feeding the commercial HD gas to the distillation unit, the distribution of the gas concentrations in the distillation unit reached to equilibrium in a few hours. 
The impurities, H$_2$ and D$_2$, were concentrated in the condenser and reboiler parts respectively.
The temperatures at the condenser and reboiler parts were 20.5 K and 23.0 K, respectively.
Fig.~\ref{fig:concentrations_before.eps}(a) shows the results of the gas analysis with the GC-QMS for the gas from the condenser part in Run 2 before the extraction.
The p-H$_2$ and o-H$_2$ peaks are dominantly observed in the spectrum with u/e=2. 
The D$_2$ component is not observed clearly.
The concentrations of the H$_2$, HD, and D$_2$ are 99.963$\pm$0.003\%, 0.035$\pm$0.001\%, and 0.002$\pm$0.001\%, respectively.
At the same location of the H$_2$ peaks, other peaks are observed in the spectrum with u/e=3.
The peaks are due to H$_3^+$ produced from the H$_2$ component as a fragmentation. 
Fig.~\ref{fig:concentrations_before.eps}(b) shows the results of the gas analysis for the gas from the reboiler part before the extraction.
The HD peak is dominantly observed in the spectrum with u/e=3. The D$_2$ peak is observed in the spectrum with u/e=4.
The concentrations of the H$_2$, HD, and D$_2$ are 0.0018$\pm$0.0002\%, 95.036$\pm$0.004\%, and 4.962$\pm$0.002\%, respectively.

An effective value of the NTP, which is an indication of the distillation performance, is derived from the concentrations of H$_2$ in the condenser and reboiler parts.
The effective value of the NTP is experimentally obtained as
\small{
\begin{equation}
\hspace{-6mm}NTP= \frac{ \displaystyle \ln \left(\frac{99.963\times10^{-2}}{1-99.963\times10^{-2}}\right) - \ln \left( \frac{0.0018\times10^{-2}}{1-0.0018\times10^{-2}}\right) }{ \ln \alpha_g}=37.2
\end{equation}
}
where $\alpha_g$ is 1.66, which  is the geometrical mean for 1.74 and 1.58, at the temperatures of 20.5 K and 23.0 K, respectively~\cite{Chem. Eng. Hand.}.
By taking the uncertainties of all the parameters into account, 37.2$\pm$0.6 is obtained for the NTP.
The result of the NTP is in good agreement with a designed value of 37.9.

\subsection{Gas analysis of purified HD}
\begin{figure}[htbp]
  \begin{center}
    \includegraphics[width=85mm]{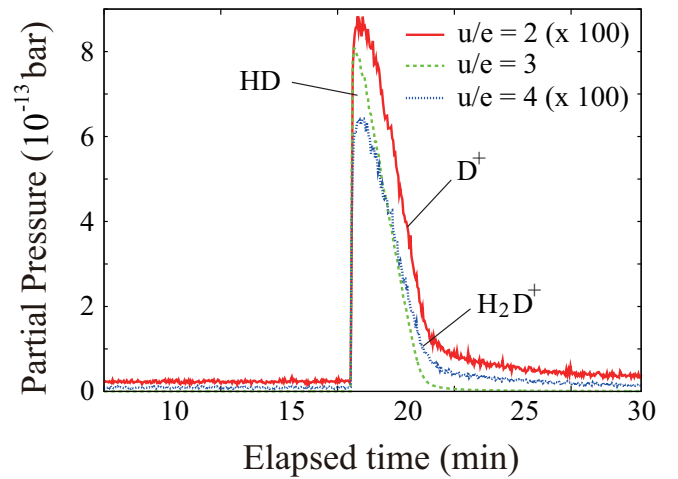}
  \end{center}
  \caption{Results of the gas analysis with the GC-QMS for the distilled gas after the 14-day gas extraction at 20.5 K, and with an extraction speed of 1.3 ml/min. Notations are the same as in Fig.~\ref{fig:concentrations_ini.eps}. }
\label{fig:concentrations_after.eps}
\end{figure}

The gas with a high H$_2$ concentration was extracted from the condenser part, and the HD concentration increased gradually.
Fig.~\ref{fig:concentrations_after.eps} shows results of the gas analysis near the end of the distillation in Run 2.
The H$_2$ and D$_2$ components are not observed at the level of 0.001\%.
The concentration of the HD component is obtained to be 99.999$\pm$0.002\%.

\subsection{Reduction of the H$_2$ concentration}
\begin{figure*}[htb]
  \begin{center}
    \includegraphics[width=150mm]{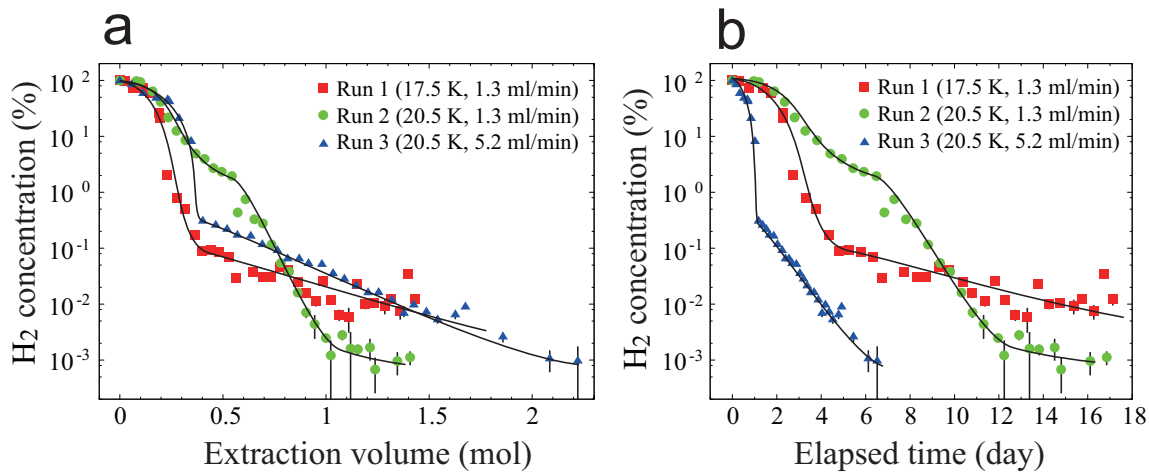}
  \end{center}
  \caption{(a) The H$_2$ concentration in the distilled gas as a function of extracted gas volume. (b) The H$_2$ concentration in the distilled gas as a function of elapsed time. The squares are the results of the concentration with a temperature of 17.5 K and an extraction speed of 1.3 ml/min (Run 1). The circles are the results of the concentration with a temperature of 20.5 K and an extraction speed of 1.3 ml/min (Run 2). The triangles are the results of the concentration with a temperature of 20.5 K and an extraction speed of 5.2 ml/min (Run 3). The curves are given for convenience to guide the eyes.}
  \label{fig:Distillation2.eps}
\end{figure*}

Fig.~\ref{fig:Distillation2.eps}(a) shows a concentration of H$_2$ including both p-H$_{2}$ and o-H$_{2}$, as a function of the extracted volume.
The concentrations of the gas extracted from the condenser part are measured for Run 1, Run 2, and Run 3.
The H$_2$ concentration at the condenser part is nearly 100\% before extracting the gas.
The H$_2$ concentration decreases as the gas is extracted from the distillation unit.
At the end of Run 2, we confirmed that a very small amount of the D$_2$ component was detected, where D$_2$ concentration was 0.008$\pm$0.001\%. 
The H$_2$ concentration reaches to 0.01\% in Run 1.
The H$_2$ concentrations reach the level under 0.001\%, which is near the detection limit of the GC-QMS, after extracting 1.0 mol and 2.0 mol gases, respectively, in Run 2 and Run 3.

The H$_2$ concentration decreases rapidly in the early step, and gradually reaches the level under 0.1\% at the extraction volume of 0.4 mol in Run 1.
The H$_2$ concentration decreases continuously in Run 2.
Judging from the comparison between Run 1 and Run 2, a small amount of the H$_2$ gas might have been trapped inside the solid D$_2$ around the condenser part at 17.5 K, and evaporates slightly in Run 1.
The extraction efficiency of the H$_2$ component at the condenser-part temperature of 20.5 K is better than that at 17.5 K. 
The H$_2$ concentration drops rapidly to 0.3\% in the early step, and gradually decreases with increasing the extraction volume in Run 3.

Fig.~\ref{fig:Distillation2.eps}(b) shows H$_2$ concentration as a function of elapsed time.
In Run 1, the H$_2$ concentration decreases rapidly in the period of 4 days, and gradually reaches the level under 0.1\% at the extraction of 17 days.
In Run 2, the H$_2$ concentration decreases continuously, and reaches to the level of 0.001\% after a long extraction time of 14.8 days.
In Run 3, the H$_2$ concentration reaches to 0.01\% at the extraction time of 4 days and 0.001\% at the extraction time of 6 day.
The periods of 17 days are needed to obtain the pure HD gas in Run 1 and Run 2, while the period is only 7 days in Run 3.
At the end, we have obtained 1 mol of HD gas with a purity better than 99.99\% from the extraction volume of 1.25 mol to 2.25 mol within a week in Run 3.
In addition, the concentration of o-H$_2$ in the produced pure HD gas is smaller than 0.001\%.
 
\section{Summary and future prospect}
We have developed a new cryogenic distillation system in order to produce a pure HD gas for the polarized HD target.
The distillation system is equipped with a cryogenic distillation unit, which is cooled at 17$\sim$21 K, filled with Heli-packs.
We succeeded in obtaining 1 mol of the HD gas with a purity better than 99.99\% for producing the polarized HD target. 
The highest concentration of HD is 99.999$\pm$0.002\% at the elapsed time of 14.8 days in Run 2.
The effective NTP, an indicator of the distillation performance, is obtained to be 37.2$\pm$0.6, which is in good agreement with a designed value of 37.9.

Since the maximum amount of the D$_2$ component was 0.008\% in the obtained high purity HD gas, the second distillation~\cite{Bouchigny09} for reducing the concentration of the D$_2$ component is not required.
In addition, we tried to shorten the operation period for the HD distillation.
The mass flow rate was set at 5.2 ml/min.
The HD distillation period was, in the present work, shortened from 30 days~\cite{Bouchigny09} to 7 days.
Since the pure HD gas can be produced with the new distillation system, the HD target will be efficiently polarized by adding an optimal amount of o-H$_2$ to the HD gas for future LEPS experiments at SPring-8.

Finally, it should be noted that during the course of the present work, we recognize that distillation using the Heli-packs is carried out in other laboratories, such as Los Alamos National Laboratory in USA for the ITER~\cite{Iwai02} and St. Petersburg Nuclear Physics Institute in Russia for the MuCap experiment~\cite{MuCap}.
The distillation system with Heli-packs for producing pure hydrogen isotopes will be more commonly used for various scientific and industrial applications in the near future.

\section{Acknowledgments}
The present work is supported in part by the Ministry of Education, Science, Sports and Culture of Japan and by the National Science Council of Republic of China (Taiwan).
This work is also supported by Program for Enhancing Systematic Education in Graduate Schools in Osaka University.
We thank Dr. T. Kageya for fruitful discussions, and are grateful to Prof. T. Kishimoto and Prof. S. Gales for their encouragements.

\end{document}